# DIRECT DETERMINATION OF THE FULLY DIFFERENTIAL CROSS SECTION OF THE IONIZATION BY THE WAVE FUNCTION IN THE COORDINATE REPRESENTATION

Zorigt Gombosuren[a], Khenmedekh Lochin[a], Aldarmaa Chuluunbaatar[a],

[a]Department of Physics, Mongolian University of Science and Technology, Ulaanbaatar city, Mongolia

This paper is focused on showing that the fully differential cross section of ionization during a collision of a proton and an antiproton with a hydrogen atom is directly expressed by the wave function in the coordinate representation. Wave function is converted from impact parameter representation to momentum transfer representation. In this case, the calculated wave function corresponds at a certain projectile angle of scattering. The square representation of this wave function module shows how the angular distribution of ionized electrons varies over time. The ionized electron's flow is directly visible from the image. In addition, the direct differentiation of the fully differential ionization cross section by the wave function is shown to be consistent with the effective time determined by the ionization amplitude, which is the traditional method.

## I. INTRODUCTION

One of the quantities measured experimentally in a result of atomic collisions with charged particles is the fully differential cross section (FDCS). In the articles of Jones and Madison [1], Voitkiv and Ullrich [2] have shown that the FDCS of the proton-hydrogen and antiproton-hydrogen collisions are calculated by the perturbation theory. Recent years non-invasive theoretical calculations have been developed by many researchers [3-10]. Although there have been experiments in measuring the differential cross-section of ionization in case of some inert gases and heavy atoms and molecules are charged by multiple lasers, the FDCS experiment has not yet been executed for the ionization of hydrogen atoms by protons and antiprotons. We calculated the electron-wave function of the hydrogen atom as a time-dependent Schrödinger equation (TDSE) as a representation of the Coulomb wave function's discrete variable (CWDVR) [11 - 13] in impact parameter method. The time-dependent probability density distribution of electron in terms of the wave function at a specific value is shown in paper [16 - 19].

aldarmaa@must.edu.mn

We converted the wave function from the representation of the impact parameter to the representation of the momentum by the Fourier transform to obtain the probability density distribution of the electrons corresponding to a specific scatter. In this paper we aim to prove that it is possible to directly determine the FDCS by the wave function in the representation of the transmitted momentum. The atomic unit system is used in this work.

## II. CONVERT THE ELECTRON WAVE FUNCTION TO A REPRESENTATION OF THE MOMENTUM TRANSFER

The collison phenomenon between charged particle (projectile) and hydrogen atom is determined by the Schrödinger equation as a time function. TDSE is written as follows.

$$i\frac{\partial}{\partial t}\Psi(\vec{r},t) = [\hat{H}_0 + \hat{V}(\vec{r},t)]\Psi(\vec{r},t) \quad (1)$$

$$\hat{V}(\vec{r},t) = \frac{-Z_p}{|\vec{R}(b,0,vt)-\vec{r}|}. \quad (2)$$

Here, $t$-is the time, $b$-is impact factor, $v$-is the velocity of the projectile, $z$- projectile charge $\vec{R} = \vec{v}t + \vec{b}$. Figure 1 shows the kinematic scheme and parameters of the collision. Denotes $\Psi_{\vec{b}}(t,\vec{r})$ the wave function denotes the correspond to $b$ -impact factor.

Using the wave functions calculated for the representation of the impact factor to obtain the wave function corresponding to a specific scattering angle or perpendicularly shifted pulse $\vec{\eta}$ for the projectile ion by the two-dimensional Fourier transform (3).

$$\Psi_{\vec{\eta}}(t,\vec{r}) = \frac{1}{2\pi} \int d\vec{b}\, e^{i\vec{\eta}\vec{b}} e^{i\delta(b)} \Psi_{\vec{b}}(t,\vec{r}) \quad (3)$$

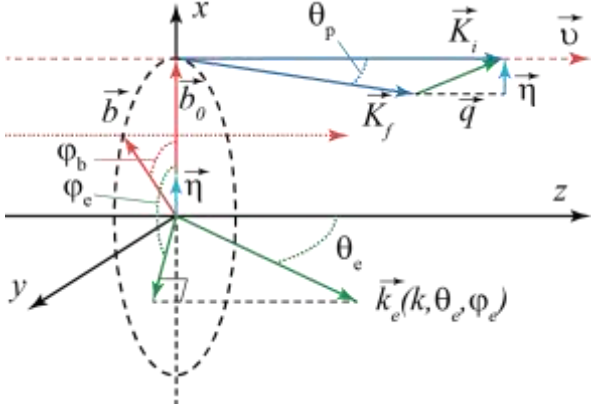

**Figure. 1**. Collision scheme of projectile ion and hydrogen atom. Projectile moves in a straight line along the z axis. $\vec{K}_i$, $\vec{K}_f$ – are the initial and final momentum of the projectile ion, $\vec{k}_e$- is the electron momentum, $\vec{\eta}$ –is (perpendicular to $\vec{v}$) component of the projectile momentum transfer $\vec{q}$

Here, $\delta(b)$ is the phase at which the nucleus of the hydrogen atoms interaction [10].

$$\delta(b) = \frac{2 Z_a Z_p}{v} \cdot \ln(v \cdot b). \quad (4)$$

Atomic nucleus and projectile charge are correspond to $Z_a = 1$ ба $Z_p = 1$ respectively. In equation (3), the wave function is defined as the coordinate representation for an electron corresponding to the specific value of the perpendicular constructor of the momentum transfer for the projectile. We consider the possibility of directly defining the FDCS by the wave function (3) in the representation of the momentum transfer.

The electron probability density is written as follows:

$$\frac{d^3 P(\vec{\eta})}{dV d\vec{\eta}} = |\Psi_{\vec{\eta}}(t,\vec{r})|^2 \quad (5)$$

In this case, we write the differential of the volume in the spherical coordinate system.

$\varepsilon$ is a function dependent on r and we write the differential probability by transferring the variable as follows

$$\frac{d^3 P(\vec{\eta})}{d\varepsilon d\Omega_e d\vec{\eta}} = \frac{|\Psi_{\vec{\eta}}(t,\vec{r})|^2 r^2 dr}{d\varepsilon}. \quad (t \to \infty) \quad (6)$$

Here, $\varepsilon$-is ejection energy. In the relativistic coordinate system [5], when the scattering angle is small, the differential of the scattering physical angle from d$\vec{\eta}$ shifts to $d\Omega_P$ and the product $K_i K_f$ will be obtained.

If the target projectile has a unit flow density, the FDCS is written as follows:

$$\frac{d^3 \sigma}{d\varepsilon d\Omega_e d\Omega_P} = K_i K_f \frac{|\Psi_{\vec{\eta}}(t,\vec{r})|^2 r^2}{d\varepsilon/dr}. \quad (t \to \infty) \quad (7)$$

Thus, to determine the relationship between the ejection energy $\varepsilon$ and the coordinates, written the wave function as follows.

$$|\Psi_{\vec{\eta}}(t,\vec{r})| = a, \quad \arg\left(\Psi_{\vec{\eta}}(t,\vec{r})\right) = S \implies \Psi = a\, e^{iS} \quad (8)$$

Here, the phase of the wave function is S and the module is $a$. This can be replaced by the Schrödinger equation to produce the following quantum Hamilton-Jacob equation [15].

$$-\frac{\partial}{\partial t} S(\vec{r},t) = \frac{1}{2}\left(\nabla S(\vec{r},t)\right)^2 + U - \frac{1}{2a}\Delta a \quad (9)$$

Since $\nabla S(\vec{r},t)$ on the right side of the equation is a particle momentum, $\frac{1}{2}\left(\nabla S(\vec{r},t)\right)^2$ is the kinetic energy, the potential energy of the field $U$ and $-\frac{1}{2a}\Delta a$ are will be Bohm potential energy. Therefore, the sum represents the total energy. Thus, the total energy corresponding to the electron at any coordinate in every time is expressed as follows.

$$\varepsilon(\vec{r},t) = -\frac{\partial}{\partial t} S(\vec{r},t) \quad (10)$$

Determine the FDCS by finding the coordinates and radial energy derivatives of the ejection energy $\varepsilon$ from expression (10) and replacing into (7). Also, the FDCS can be defined by traditional method using amplitude in quantum mechanics. It is expressed as follows [10].

$$\frac{d^3\sigma}{d\varepsilon d\Omega_e d\Omega_P} = K_i K_f |T(\varepsilon, \theta_e, \varphi_e, \eta, \varphi_\eta)|^2 \quad (11)$$

Here, ionization amplitude is as follows:

$$T(\varepsilon, \theta_e, \varphi_e, \eta, \varphi_\eta) = \langle \Psi_{\vec{k}}^{(-)} | \Psi_{\vec{\eta}}(t,\vec{r}) \rangle. \quad (t \to \infty) \quad (12)$$

This amplitude expression (12) is similar to the results described by other researchers [10], in which the ionization amplitude is determined by the representation of the impact factor and converted to the representation of transfer momentum through the Fourier transform. This is projected on $\Psi_{\vec{k}}^{(-)}$ and is analytically verified by changing the order of the Fourier transform integrals. The new expression FDCS (7) defined by wave function is compared to the FDCS calculated by traditional method (11).

## III. CALCULATION RESULTS

CWDVR calculation were performed by the T DSE when energy of the antiproton and proton at 200 keV. In this method, the projectile electron interaction is represented as a time-dependent field in the representation of the impact factor, the electron wave function is decomposed into a spherical harmonic function in a spherical coordinate system, and the radial function is calculated as a time-dependent function. The radial coordinate were discreted by the roots of the Coulomb wave function and construction of the pseudo spectral bases of the hydrogen atoms were calculated. The parameters are taken as follows in the calculation [13]. For antiprotons, the charge number of the Coulomb wave function was Z = 120, the number of waves was k = 8, and the radial node was N = 1200, while r $_{max}$ = 457.8 . The maximum value of the orbital quantum number was taken as 8 and the calculation was calculated by shifting the impact factor $b_{max}$ = 30.8 and the initial position of projectile ion z is from -80 to 1000 in steps of Δz = 0.08. For protons, the values of the parameters are Z = 120, k = 4, N = 800, $r_{max}$ = 583.1, $b_{max}$ = 37.5, z = 560, and Δz = 0.16 respectively.

In the theoretical works on the process of collisions of protons and antiprotons with hydrogen atoms, the evolution of the wave function at the particular impact factor values is described by the square of its modulus [16-19]. Although this represents the evolution of the collision process, it cannot be measured by experiment. This is because it is not yet possible to conduct experiments and measurements at the particular impact factor. We described the wave function through equation (3) to represent the transfer momentum to show experimentally measurable results. This wave function allows us to see the evolution of the collision while representing quantities that can be measured experimentally. Figure 2 and 3 show the square of the wave function modules ($|\Psi_{\vec{\eta}}(t,\vec{r})|^2 r^2$) at the scattering plane at the scattering angle of protons and antiprotons 0.0002 rad or perpendicular component of transfer momentum η = 0.519. It illustrates the antiproton z coordinates 160, 2400, 320 and the proton z coordinates 240, 320 and 400 at different moments. The moving outwards of ionized electrons represented by peak shift is shown in Figure 2.

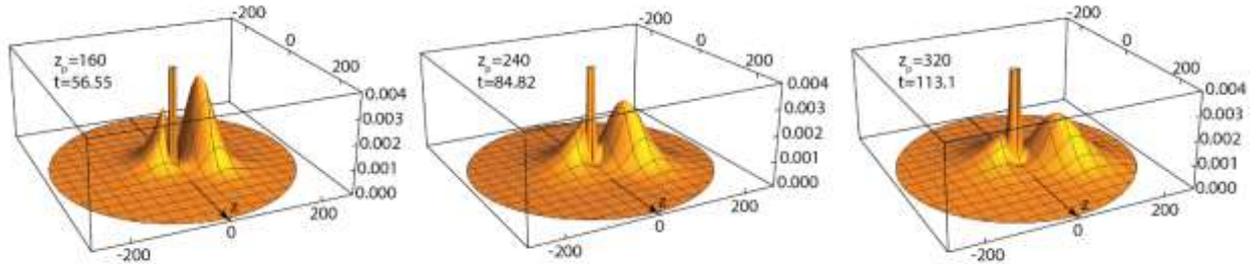

Figure 2. z coordinate $Z_p$ of antiproton and time t , electron probability density in the scattering plane

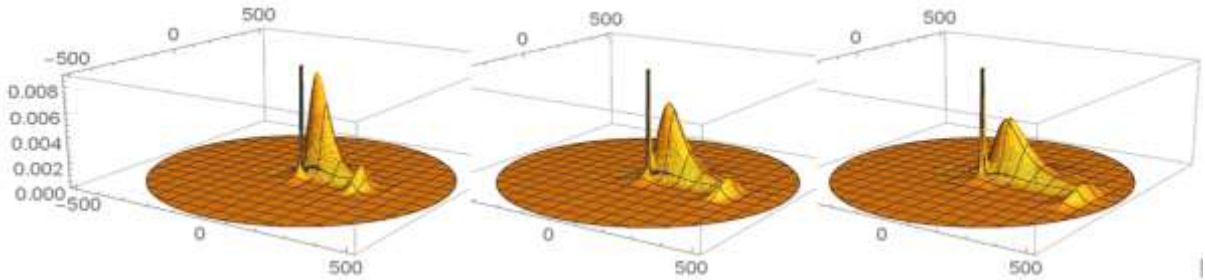

**Figure 3.** The probability density of electrons in the corresponding scattering plane when the z-coordinate of the proton is 240, 320, 400.

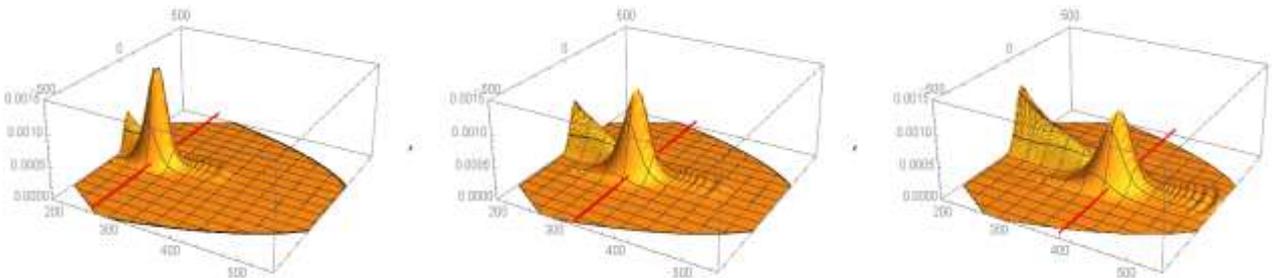

*Figure 4. The probability density of an electron in the plane of scattering corresponding to the moment of time when the z-coordinate of the proton is 272,336,400 (represented by red line).*

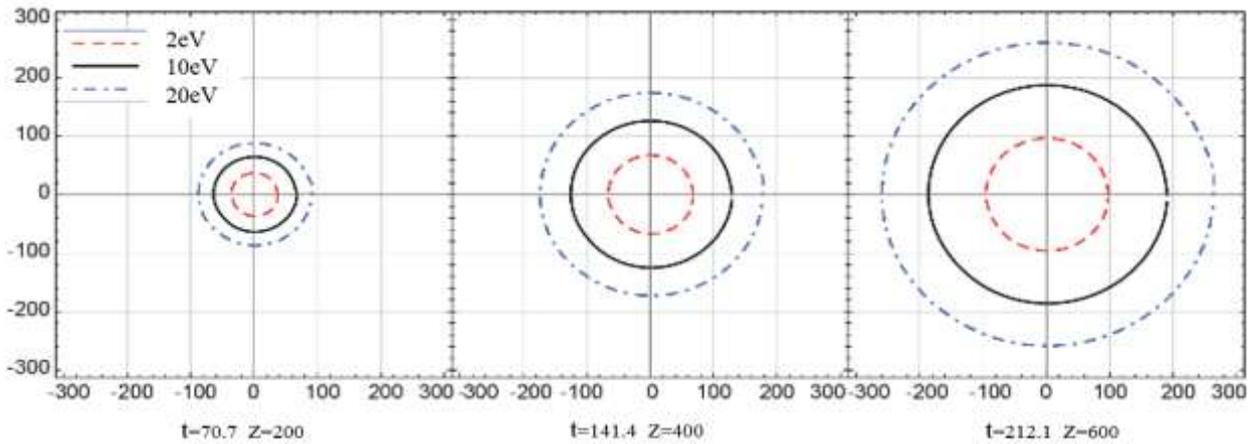

*Figure 6. Time evolution of the same energy surface on the scattering plane. Here t is the time and zp is the z coordinate of the antiproton.*

Proton collisions cause excitation, ionization, and charge transfer. In Figure 3, the transferred charge with proton is shown as a small peak shifting along the z axis. To verify this process, position of the proton at each moment of time with the z coordinates with values of 272, 336, and 400 compares the low-peak place. As shown in Figure 4, the low-peak place coincides with the position of the proton, indicating that the electron is moving together with the proton in this region. The dependence of the energy coordinates from the numerical solution of the TDSE is determined by Equation (10). Figure 5 shows the dependence of the energy coordinates in the direction perpendicular to the z-axis on the plane of scattering. The increase in energy depending on the coordinates indicates that the high-energy electrons are moving away from the center of the atom first. Also, as time increases, the coordinates corresponding to the same energy increase, indicating that the electrons are moving outwards.

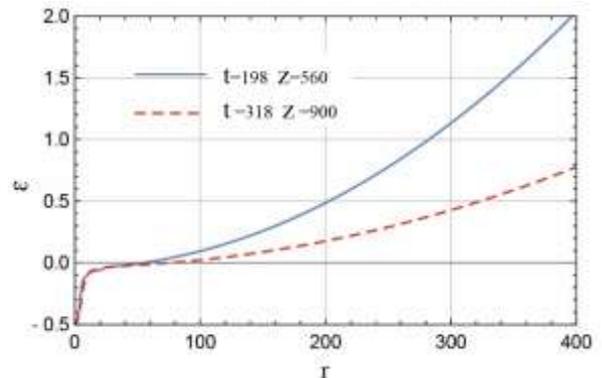

*Figure. 5. Coordinate energy dependence in 2 time moment. Direction of radial coordinate is perpendicular velocity and on the scattering plane*

As shown in Figure 6, the energy surface is distributed in a radial direction, almost in the spherical form. The deviation of the energy surface from spherical form surface shall not exceed 3%. The energy coordinate relationship was found in all directions at each moment of time, and the FDCS was calculated as a wave function on the surface corresponding to a specific energy. As shown in Figure 7, the FDCS, which was determined by our numerical results using TDSE, coincides with the results calculated by the relativistic method [10] and calculated by the quantum mechanical method [7].

It has also previously shown that the results calculated by the CWDVR method in case of other energy are consistent with other theoretical calculations [13]. Therefore, the new expression FDCS (7) defined by wave function is compared to the FDCS calculated by traditional method (11).

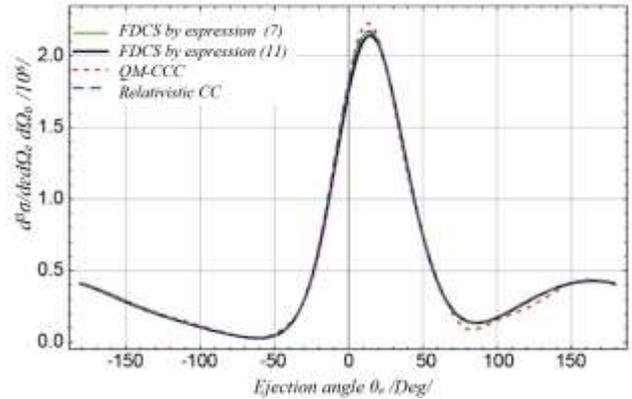

*Figure 7. When an antiproton with an energy of 200 keV is corresponding to scatering angle of 0.2 mrad, an electron FDCS ionized with an energy of 7 eV is plotted on the scattering plane. Here horizontal axis is angle $\theta_e$. Green line – FDCS by espression (7), black line – FDCS by espression (11), blue dashed line– relativistic CC [10], dots - QM-CCC [7].*

For the proton-hydrogen atomic collision, the wave function (3) is used to find the FDCS by Equation (7) and comparison with the traditional method (11) is shown in Figure 8.

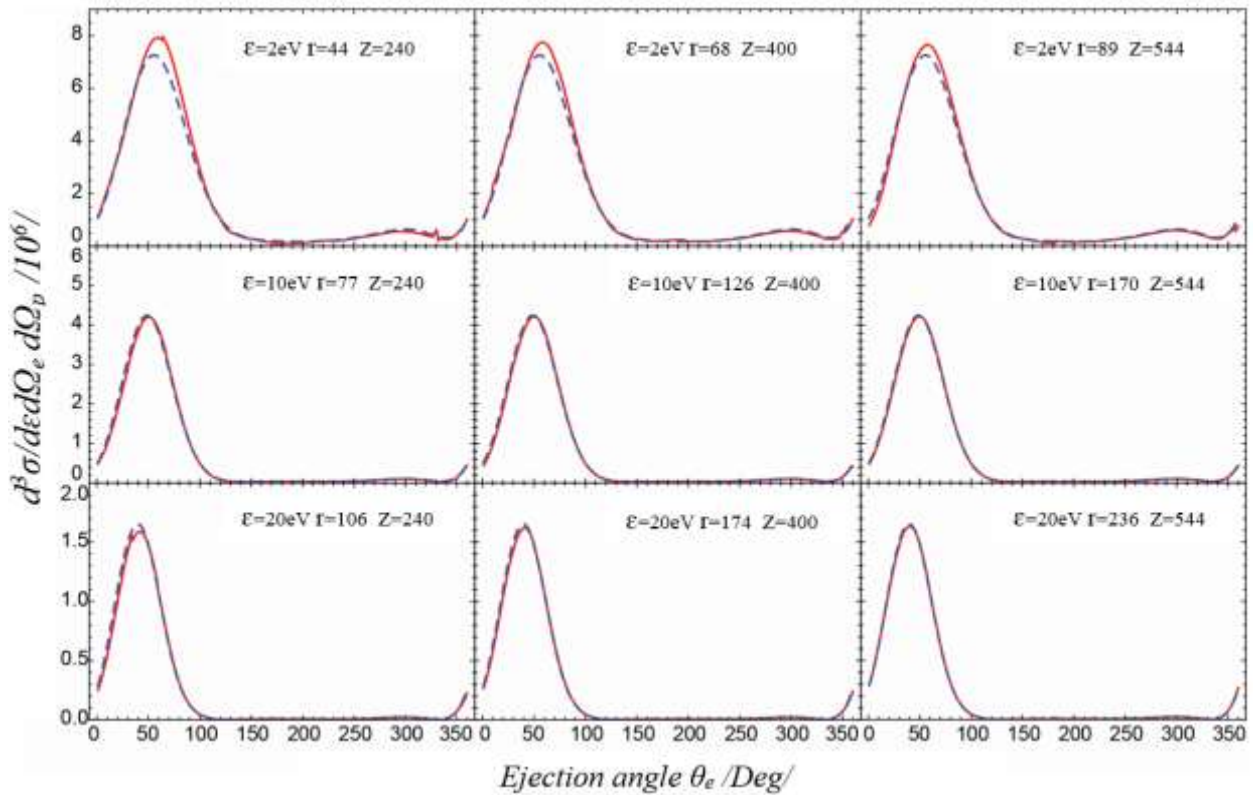

*Figure 8. The FDCS illustrates at three different moments of the proton z coordinates 240, 400, and 544 in the projection plane when a proton scattering angle of 0.2 mrad or a perpendicular constructor of the transfer momentum is $\eta = 0.519$. Here ε is the ejection energy, z is the z-coordinate of the proton, and r is the radial coordinate of the electron. The closed FDCS red curve shown in the figure is FDCS by new expression (7) and the blue dashed curve is FDCS by traditional method (11).*

The coincidence of these results was assessed. Assuming a relative difference in the maximum value of the FDCS, the ejection energy is decreasing from 3.9% to 2.4% at 2 eV, from 0.9% to 0.6% at 10 eV, and from 2.8% to 0.8% at 20 eV when the proton position is between 240 and 544. This indicates that as the ionized electron moves away from the atomic nucleus, the FDCS by new expression (7) corresponds to the FDCS by traditional method (11).

The calculated redults of antiproton- hydrogen atomic collision is shown in Figure 9.

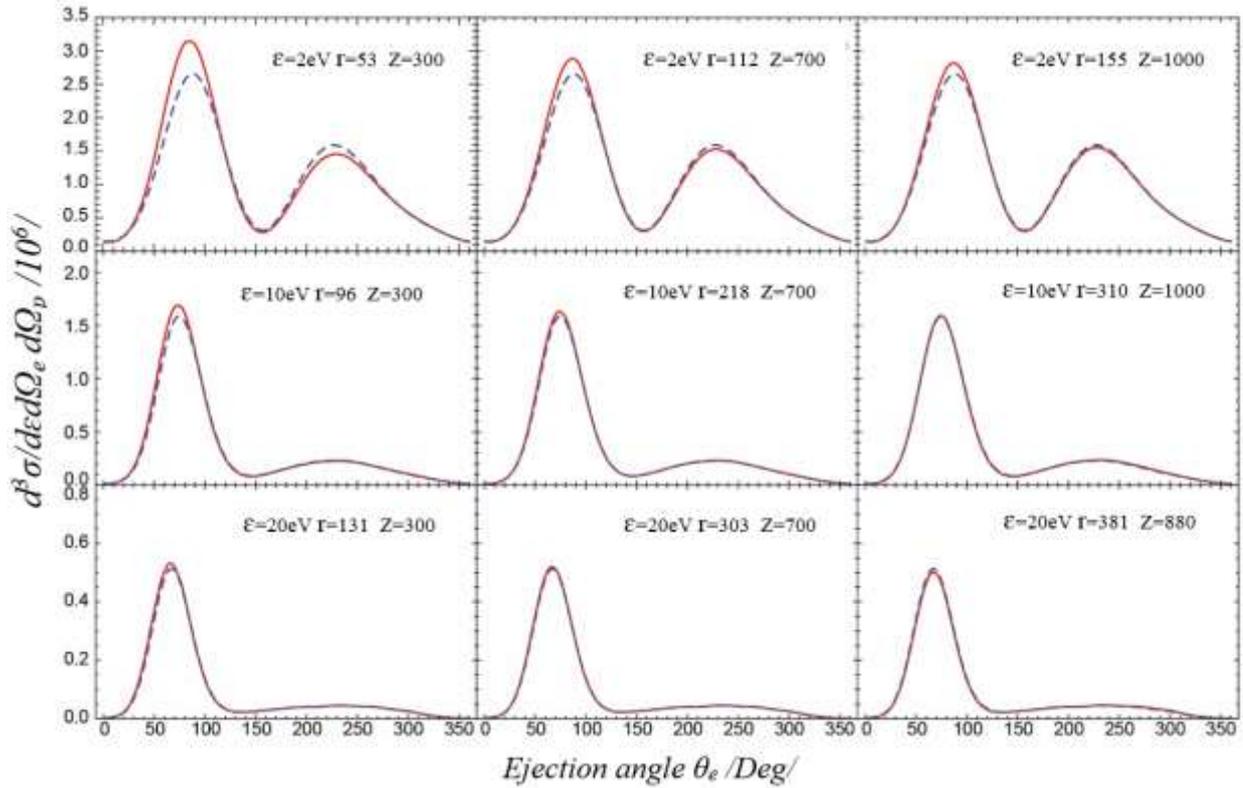

*Figure 9. The FDCS illustrates at three different moments of the proton z coordinates in the projection plane when a antiproton scattering angle of 0.2 mrad. Here ε is the ejection energy, z is the z-coordinate of the antiproton, and r is the radial coordinate of the electron. The closed FDCS red curve shown in the figure is FDCS by new expression (7) and the blue dashed curve is FDCS by traditional method (11).*

As can be seen in Figure 8 and 9, FDCS results in above mentioned 2 method, requires a long time to converge in 2eV ejection energy. It can be explained the cause that, electrons with low energy get away slowly from nuclei. Bound state can be affect in FDCS results. On the other hand, when ejection energy increases, FDCS results rapidly converges is depicted in Figure 8 and 9.

**IV. CONCLUSIONS**

The result of FDCS (7) calculated by time-dependent wave function is agreement with the value of the FDCS (11) calculated by amplitude in a large value of time and electron moves away from the nucleus. This method has the disadvantage that, it needs to be calculated over a large range of space and time compared to the traditional method, but advantage is it does not use the continuum wave function. For multi-electron atoms, continuum wave function is not readily defined, making it difficult to calculate the FDCS. This problem can be avoided by using the new expression FDCS (7).

**V. REFERENCES**
[1] S. Jones and D. H. Madison, Phys. Rev. A 65, (2002) 052727.
[2] A. B. Voitkiv and J. Ullrich, Phys. Rev. A 67, (2003) 062703.
[3] A. Igarashi, S. Nakazaki, and A. Ohsaki, Phys. Rev. A, Vol 61, (2000) 062712.
[4] Xiao-Min Tong, Tsutomu Watanabe, Daiji Kato and Shunsuke Ohtani. Phys. Rev. A, volume 64, (2001) 022711.
[5] M. McGovern, D. Assafrao, J. R. Mohallem, C. T. Whelan, and H. R. J. Walters, Phys. Rev. A 79, (2009) 042707.
[6] M. McGovern, D. Assafrao, J. R. Mohallem, C. T. Whelan, and H. R. J. Walters, Phys. Rev. A 81, (2010) 032708.
[7] I. B. Abdurakhmanov, A. S. Kadyrov, I. Bray, and A. T. Stelbovics, J. Phys. B: At. Mol. Opt. Phys. 44, (2011) 165203.
[8] M. F. Ciappina, T.G. Lee, M. S. Pindzola, and J. Colgan, Phys. Rev. A 88, (2013) 042714.
[9] I. B. Abdurakhmanov, A. S. Kadyrov, and I. Bray, Phys. Rev. A 94, (2016) 022703.
[10] A. I. Bondarev, Y. S. Kozhedub, I. I. Tupitsyn, V. M. Shabaev, and G. Plunien. Phys. Rev. A 95, (2017) 052709.
[11] K.M. Dunseath, J.M Launay, M Terao-Dunseath and L Mouret J. Phys. B: At. Mol. Opt. Phys. 35 (2002) 3539–3556
[12] Peng. Liang-You and Starace. Anthony F, The journal of chemical physics 125, (2006) 154311.
[13] G. Zorigt, L. Khenmedekh, Ch. Aldarmaa, IJMA- 10(5), (2019) 19-23.
[14] A. I. Bondarev, I. I. Tupitsyn, I.A.Maltsev,Y. S. Kozhedub,, and G. Plunien. Eur.Phys. J.D 69,110 (2015).

[15] L. D. Landay and E. M. Lifshitz, Quantum Mechanics. Norelativistic Theory, 4th ed . ( 1989).
[16] J. Azuma, N. Toshima, K. Hino, and A. Igarashi, Phys. Rev. A 64, 062704 (2001).
[17] Emil Y Sidky and C D Lin. J. Phys. B: At. Mol. Opt. Phys. 31 (1998) 2949–2960
[18] J. C. Wells, D. R. Schultz, P. Gavras and M. S. Pindzola. Phys. Rev. A 54, (1996)
**[19]** B. Pons. Phys. Rev. A 63, (2000) 0127